          [2004/06/22 1.2 (KOF); 2001/04/25 1.1 (PWD)]

\documentclass[a4paper,twocolumn]{Gaia2004} 
\usepackage{times}      
\usepackage{epsfig}     
\usepackage{natbib}     
\title{[$\alpha$/Fe] in the thin and the thick disk -
towards an automatic parametrization of stellar spectra.}

\author{Pascal Girard, }
\author{Caroline Soubiran}
\affil{Observatoire Aquitain des Sciences de l'Univers, L3AB, 
2 rue de l'Observatoire, BP 89, 33270 Floirac, France.}

\bibpunct{(}{)}{;}{a}{}{,}  

\begin{document}

\keywords{methods: data analysis - stars: abundances - 
stars: fundamental parameters}

\maketitle

\begin{abstract}
We test an automatic procedure to measure [Fe/H] and 
[$\alpha$/Fe] on high resolution spectra. The test sample is the intersection of 
the ELODIE library and a catalogue of 830 stars having well determined abundances.

\end{abstract}

\section{Goal}
In order to investigate the properties of the thick disk and its interface 
with the thin disk we have compiled a catalogue of elemental abundances 
of O, Na, Mg, Al, Si, Ca, Ti, Ni, Fe including 830 stars (Girard and Soubiran 2004). 
The classification of 
thin disk and thick disks stars has been performed on the basis of their 
(U,V,W) velocities. The two populations overlap greatly in metallicity but 
at a given [Fe/H] the thick disk shows on average an enhancement of 0.07 dex in 
[$\alpha$/Fe] (Fig.1). In order to go further in this investigation we want to be able 
to measure [Fe/H] and [$\alpha$/Fe] from a large collection of spectra with 
an automatic procedure.

 \begin{figure}[h]
  \begin{center}
    \leavevmode
 \centerline{\epsfig{file=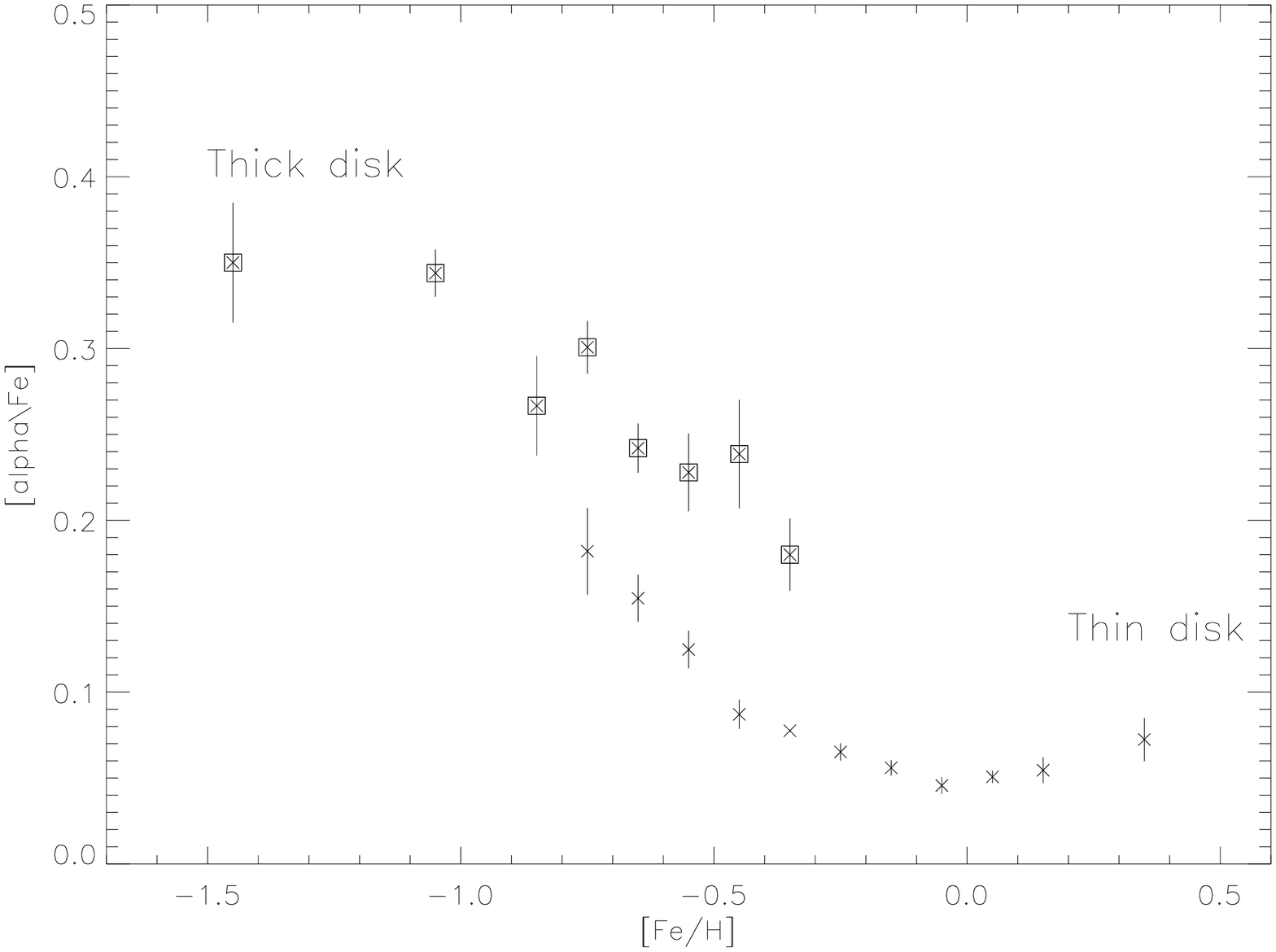,angle=90, width=8.0cm}}
   \end{center}
  \caption{[$\alpha$/Fe] vs [Fe/H]}
  \label{}
\end{figure}
 \begin{figure}[h!]
  \begin{center}
    \leavevmode
 \centerline{\epsfig{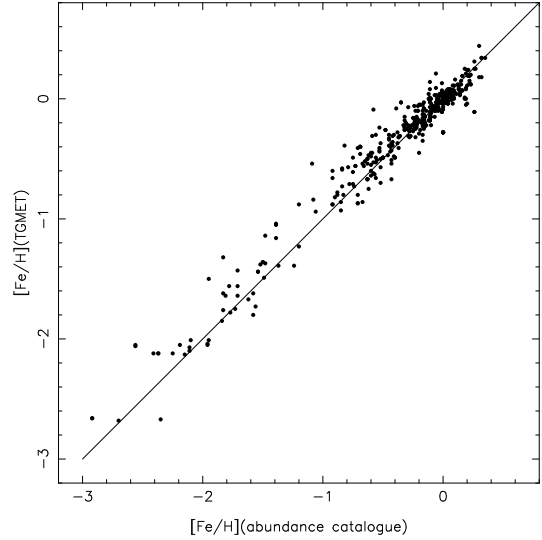}}
   \end{center}
  \caption{Comparison of [Fe/H] from the TGMET code with 
  [Fe/H] from the catalogue of abundances.  Rms = 0.11.}
  \label{}
\end{figure}

\section{Tools and material}
In this section we summerize the libraries and the codes 
used for this investigation : 

\begin{itemize}
\item The ELODIE library of 1962 spectra ($\lambda \lambda$390.6-681.1nm, R=42000) 
of 1388 stars with measured Lick indices 
(Prugniel \& Soubiran 2004) and its intersection with
the abundance catalogue : 449 spectra of 308 stars.
\item The grid of synthetic spectra with 3 values of [$\alpha$/Fe] 
(Barbuy et al. 2003).
\item The TGMET code : a minimun 
distance algorithm to measure (Teff, logg, [Fe/H]) (Katz et al. 1998).
\item The ETOILE code : a modified version of TGMET 
with determination of [$\alpha$/Fe] (D.Katz, priv. com.).
\\
\\
\end{itemize}

\begin{figure}[h]
  \begin{center}
    \leavevmode
 \centerline{\epsfig{file=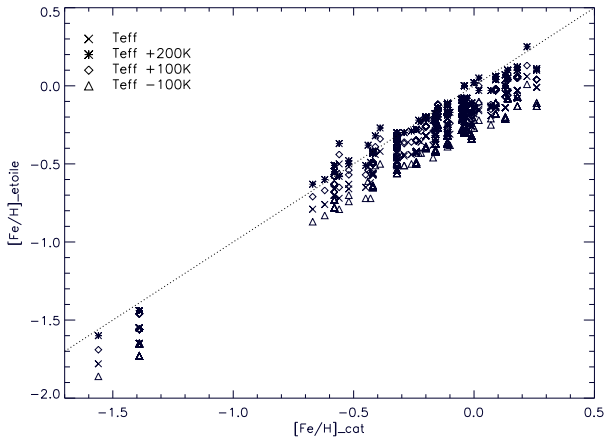,angle=90, width=8.0cm}}
   \end{center}
  \caption{[Fe/H] from ETOILE vs [Fe/H] from the catalogue.
  The modification of Teff in the input of the code provides a variation of [Fe/H].}
  \label{}
\end{figure}

 \begin{figure}[h!]
  \begin{center}
    \leavevmode
 \centerline{\epsfig{file=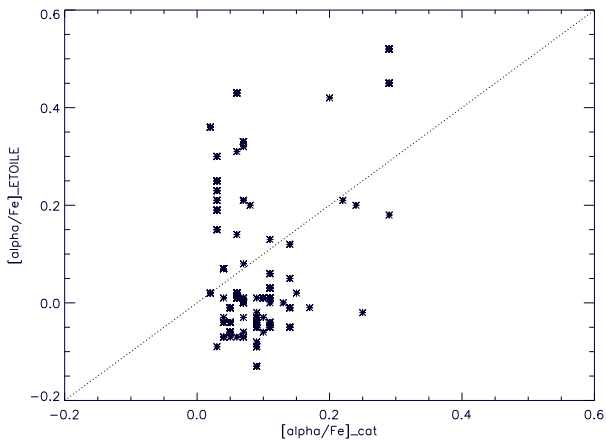,angle=90, width=8.0cm}}
   \end{center}
  \caption{[$\alpha$/Fe] from ETOILE vs [$\alpha$/Fe] from the catalogue.}
  \label{}
\end{figure}
 
TGMET relies on the least-square comparison of an ELODIE 
spectrum of a target star to a library of ELODIE spectra of reference 
stars with well determined atmospheric 
parameters.\\
ETOILE is a minimum distance algorithm based on the perturbation 
method described in Cayrel et al. (1991). With this method, 
the reference library must sample the parameter space with 
regular steps. 
That is why synthetic spectra are used instead of empirical 
spectra.\\
We use the grid of synthetic spectra computed by Barbuy et al.(2003) : 
$\lambda \lambda$460-560nm, 4000 $\leq$ Teff $\leq$ 7000 K in steps of 250 K,
0.5 $\leq$ log g $\leq$ 5.0 in steps of 0.5, 
[Fe/H] : -3.0, -2.5, -2.0, -1.5, -1.0, -0.5, -0.3, -0.2, -0.1, 0.0 and +0.3 and 
[$\alpha$/Fe] : 0.0, +0.2 and +0.4. \\
A first step is to validate the grid, 
that is verify that 
computed spectra and observed ones with same parameters match 
on the whole wavelength interval.

\section{Results}
A bootstrap method is used to test the performances of TGMET. 
Based on 449 spectra, TGMET is able to 
retrieve the atmospheric parameters 
with a typical accuracy of 134K in Teff and 0.11 in 
[Fe/H] (Fig.2). 
The main limitation of TGMET is its empirical reference library 
which does not sample perfectly the parameter space. A limitation 
overcome with the use of ETOILE and a grid of synthetic spectra.

As a starting point ETOILE uses the TGMET solution.
Preliminary results from ETOILE
suggest that the catalogue of abundances and the grid are not on 
the same temperature scale : metallicities are correctly recovered 
if a hotter temperature is given in input (Fig.3). [$\alpha$/Fe] is not yet 
correctly estimated (Fig.4). Possible causes are currently investigated.

\end{document}